\documentclass[12pt]{article}

\usepackage{amsmath,amssymb,graphicx,color,cite}
\numberwithin{equation}{section}

\evensidemargin \oddsidemargin

\begin{document}

\begin{titlepage}

\renewcommand{\thefootnote}{\fnsymbol{footnote}}

\hfill\parbox{4cm}{hep-th/0309258 \\ KEK-TH-917}

\vspace{15mm}
\baselineskip 9mm
\begin{center}
  {\Large \bf One-Loop Flatness \\ 
    of Membrane Fuzzy Sphere Interaction\\
    in Plane-Wave Matrix Model}
\end{center}

\baselineskip 6mm
\vspace{10mm}
\begin{center}
  Hyeonjoon Shin$^{a,b}$\footnote{\tt hshin@newton.skku.ac.kr,
    hshin@post.kek.jp} 
  and Kentaroh Yoshida$^b$\footnote{\tt
    kyoshida@post.kek.jp} \\[10mm] 
  {\sl $^a$BK 21 Physics Research
    Division and Institute of Basic Science\\ 
    Sungkyunkwan University,
    Suwon 440-746, South Korea \\[3mm] 
    $^b$Theory Division, High Energy
    Accelerator Research Organization (KEK)\\ 
    Tsukuba, Ibaraki 305-0801, Japan}
\end{center}

\thispagestyle{empty}

\vfill
\begin{center}
{\bf Abstract}
\end{center}
\noindent
In the plane-wave matrix model, the background configuration of two
membrane fuzzy spheres, one of which rotates around the other one in
the $SO(6)$ symmetric space, is allowed as a classical solution.  We
study the one-loop quantum corrections to this background in the path
integral formulation.  Firstly, we show that each fuzzy sphere is
stable under the quantum correction.  Secondly, the effective
potential describing the interaction between fuzzy spheres is obtained
as a function of $r$, which is the distance between two fuzzy spheres.
It is shown that the effective potential is flat and hence the fuzzy
spheres do not feel any force.  The possibility on the existence of
flat directions is discussed.  
\\ [5mm] Keywords : pp-wave, Matrix model, Fuzzy sphere 
\\ PACS numbers : 11.25.-w, 11.27.+d, 12.60.Jv

\vspace{5mm}
\end{titlepage}

\baselineskip 6.6mm
\renewcommand{\thefootnote}{\arabic{footnote}}
\setcounter{footnote}{0}

\section{Introduction}

The plane-wave matrix model \cite{Berenstein:2002jq} is a microscopic
description of the discrete light cone quantized (DLCQ) M-theory in
the eleven-dimensional $pp$-wave or plane-wave background.  The
eleven-dimensional plane-wave \cite{Kowalski-Glikman:1984wv} is
maximally supersymmetric and the limiting case of the
eleven-dimensional $AdS$ type geometries \cite{Blau:2002dy}.  Its
explicit form is given by
\begin{align}
ds^2 &= -2 dx^+ dx^- 
    - \left( \sum^3_{i=1} \left( \frac{\mu}{3} \right)^2 (x^i)^2
            +\sum^9_{a=4} \left( \frac{\mu}{6} \right)^2 (x^a)^2
      \right) (dx^+)^2
    + \sum^9_{I=1} (dx^I)^2 ~,
     \notag \\
F_{+123} &= \mu ~,
\label{pp}
\end{align}
where $I=(i,a)$.  Due to the effect of the $++$ component of the
metric and the presence of the four-form field strength, the
plane-wave matrix model has some $\mu$ dependent terms, which make the
difference between the usual flat space matrix model and the
plane-wave one.

The presence of the $\mu$ dependent terms makes the plane-wave matrix
model have some peculiar properties.  One of them is that there are
various vacuum structures classified by the $SU(2)$ algebra
\cite{Dasgupta:2002hx}.  The crucial ingredient for vacua is the
membrane fuzzy sphere.  It preserves the full 16 supersymmetries of
the plane-wave matrix model and exists even at finite $N$ which is the
size of matrix.

After the plane-wave matrix model was proposed and its basic aspects
were uncovered, there have been lots of investigations in various
directions.  The structure of vacua has been studied in more detail
especially related to the protected multiplet
\cite{Dasgupta:2002hx,Kim:2002if,Dasgupta:2002ru}.  The possible BPS
objects contained in the plane-wave matrix model have been searched
\cite{Bak:2002rq,Sugiyama:2002rs,Hyun:2002cm,Sugiyama:2002jq,
Alishahiha:2002sy,Sugiyama:2002bw,Mikhailov:2002wx,Park:2002cb,
Maldacena:2002rb,Yee:2003ge,Sakaguchi:2003ah}.  The algebraic and
structural study of the model itself and the various BPS objects
present in it has been performed
\cite{Dasgupta:2002hx,Sugiyama:2002rs,Hyun:2002cm,Kim:2002zg,
Kim:2002cr,Lee:2002vx,Nakayama:2002tb,Hyun:2002fk,Hyun:2003se,
Freedman:2003kb,Kim:2003rz}.  Based on the fact that the low energy
description of the M theory is the eleven dimensional supergravity,
there also have been supergravity side analysis
\cite{Fernando:2002wv,Mas:2003uk,Kimura:2003um}.

If the M theory is compactified on a circle, then we have
ten-dimensional Type IIA string theory.  Under the circle
compactification, the $pp$-wave geometry (\ref{pp}) becomes the IIA
$pp$-wave background which is not maximally supersymmetric and has 24
supersymmetries
\cite{Michelson:2002wa,Bena:2002kq,Sugiyama:2002tf,Hyun:2002wu}.  For
the purpose of understanding the plane-wave matrix model as well as
the string theory itself, the IIA string theory in the $pp$-wave
background has been also extensively studied, in parallel with the
progress in the study of the plane-wave matrix model
\cite{Sugiyama:2002tf,Hyun:2002wu,Bonelli:2002mb,Hyun:2002wp,
  Iizuka:2002ra,Hyun:2002xe,Kwon:2003in,Hyun:2003ks,Shin:2003ae,
  Das:2003yq,Kim:2003zc}.

However, despite of quite amount of progress in the study of the
plane-wave matrix model, there has been lack of the investigation
about the dynamical aspects. However, see for example
\cite{Lee:2003kf}. In fact, the present status of the plane-wave
matrix model enables us to study the dynamics of the model.  In this
paper, we consider the basic objects of the plane-wave matrix model
and study their interaction.

For the interacting objects, we take two membrane fuzzy spheres, both
of which are supersymmetric.  In the $SO(3)$ symmetric subspace which
one may see in the $pp$-wave background (\ref{pp}), two fuzzy spheres
are taken to be at the origin.  In the $SO(6)$ symmetric space, one
fuzzy sphere is located at the origin, while the other fuzzy sphere is
taken to rotate around the origin with a fixed distance.  It should be
noted that this configuration is allowed as a classical solution of
the equations of motion and furthermore the rotating fuzzy sphere
itself is supersymmetric.  We will evaluate the one-loop corrections
to the configuration and obtain the effective potential.  As we will
see, the effective potential is flat.  This implies that the whole
configuration of fuzzy spheres is also supersymmetric.  One may argue
that the flat potential is natural since each fuzzy sphere
configuration is supersymmetric.  However, the situation is
unconventional from the viewpoint of the flat space matrix model
\cite{Banks:1997vh} and indicates one of intriguing properties of the
plane-wave matrix model.  Moreover, the flat potential shows us the
possibility that the plane-wave matrix model has the flat directions
which have not been observed in it.

The organization of this paper is as follows.  In the next section, we
give the action of the plane-wave matrix model and consider its
classical solutions focused on our concern.  The expansion of the
action around a given arbitrary background is given in section
\ref{bg-exp}.  In section \ref{fuzzy-config}, we set up the background
configuration and consider the fluctuations around it.  In section
\ref{stable}, the one-loop stability of each fuzzy sphere is checked
for arbitrary size.  In section \ref{interact}, we evaluate the path
integration of fluctuations responsible for the interaction between
fuzzy spheres.  It will be shown that the one-loop effective potential
is flat.  Thus, the fuzzy spheres do not feel any force.  Finally,
conclusion and discussion will be given in section \ref{conc}.  We
discuss the possibility on the existence of flat directions.

\section{Plane-wave matrix model and classical solutions}

The plane-wave matrix model is basically composed of two parts.  One
part is the usual matrix model based on  eleven-dimensional flat
space-time, that is, the flat space matrix model, and another is a set
of terms reflecting the structure of the maximally supersymmetric
eleven dimensional plane-wave background, Eq. (\ref{pp}).  Its action
is
\begin{equation}
S_{pp} = S_\mathrm{flat} + S_\mu ~,
\label{pp-bmn}
\end{equation}
where each part of the action on the right hand side is given by
\begin{align}
S_\mathrm{flat} & = \int dt \mathrm{Tr} 
\left( \frac{1}{2R} D_t X^I D_t X^I + \frac{R}{4} ( [ X^I, X^J] )^2
      + i \Theta^\dagger D_t \Theta 
      - R \Theta^\dagger \gamma^I [ \Theta, X^I ]
\right) ~,
  \notag \\
S_\mu &= \int dt \mathrm{Tr}
\left( 
      -\frac{1}{2R} \left( \frac{\mu}{3} \right)^2 (X^i)^2
      -\frac{1}{2R} \left( \frac{\mu}{6} \right)^2 (X^a)^2
      - i \frac{\mu}{3} \epsilon^{ijk} X^i X^j X^k
      - i \frac{\mu}{4} \Theta^\dagger \gamma^{123} \Theta
\right) ~.
\label{o-action}
\end{align}
Here, $R$ is the radius of circle compactification along $x^-$ and
$D_t$ is the covariant derivative with the gauge field $A$,
\begin{equation}
D_t = \partial_t - i [A, \: ] ~.
\end{equation}

For dealing with the problem in this paper, it is convenient to
rescale the gauge field and parameters as
\begin{equation}
A \rightarrow R A ~,~~~ 
t \rightarrow \frac{1}{R} t ~,~~~
\mu \rightarrow R \mu ~.
\end{equation}
With this rescaling, the radius parameter $R$ disappears and the
actions in Eq.~(\ref{o-action}) become
\begin{align}
S_\mathrm{flat} & = \int dt \mathrm{Tr} 
\left( \frac{1}{2} D_t X^I D_t X^I + \frac{1}{4} ( [ X^I, X^J] )^2
      + i \Theta^\dagger D_t \Theta 
      -  \Theta^\dagger \gamma^I [ \Theta, X^I ]
\right) ~,
  \notag \\
S_\mu &= \int dt \mathrm{Tr}
\left( 
      -\frac{1}{2} \left( \frac{\mu}{3} \right)^2 (X^i)^2
      -\frac{1}{2} \left( \frac{\mu}{6} \right)^2 (X^a)^2
      - i \frac{\mu}{3} \epsilon^{ijk} X^i X^j X^k
      - i \frac{\mu}{4} \Theta^\dagger \gamma^{123} \Theta
\right) ~.
\label{pp-action}
\end{align}

The possible backgrounds allowed by the plane-wave matrix model are
the classical solutions of the equations of motion for the matrix
fields.  Since the background that we are concerned about is purely
bosonic, we concentrate on solutions of the bosonic fields $X^I$.  We
would like to note that we will not consider all possible solutions
but only those relevant to our interest for the fuzzy sphere
interaction.  Then, from the rescaled action, (\ref{pp-action}), the
bosonic equations of motion are derived as
\begin{align}
\ddot{X}^i &=
   - [[ X^i, X^I],X^I] - \left( \frac{\mu}{3} \right)^2 X^i
   - i \mu \epsilon^{ijk} X^j X^k ~, 
                 \nonumber \\
\ddot{X}^a &=
    - [[ X^a, X^I],X^I] - \left( \frac{\mu}{6} \right)^2 X^a ~,
\label{eom}
\end{align}
where the over dot implies the time derivative $\partial_t$.  

Except for the trivial $X^I=0$ solution, the simplest one is the
simple harmonic oscillator solution;
\begin{equation}
X^i_\mathrm{osc} 
= A^i \cos \left( \frac{\mu}{3} t + \phi_i \right) 
 {\bf 1}_{N \times N} ~,~~~
X^a_\mathrm{osc} 
= A^a \cos \left( \frac{\mu}{6} t + \phi_a \right) 
 {\bf 1}_{N \times N} ~,
\label{osc}
\end{equation}
where $A^I$ and $\phi_I $ $(I=(i,a))$ are the amplitudes and phases of
oscillations respectively, and ${\bf 1}_{N \times N}$ is the $N \times
N$ unit matrix.  This oscillatory solution is special to the
plane-wave matrix model due to the presence of mass terms for $X^I$.
It should be noted that, because of the mass terms, the configuration
corresponding to the time dependent straight line motion, say $v^I t +
c^I$ with non-zero constants $v^I$ and $c^I$, is not possible as a
solution of (\ref{eom}), that is, a classical background of plane-wave
matrix model, contrary to the case of the flat space matrix model.  As
the generalization of the oscillatory solution, Eq. (\ref{osc}), we
get the solution of the form of diagonal matrix with each diagonal
element having independent amplitude and phase.

As for the non-trivial constant matrix solution, Eq. (\ref{eom})
allows the following membrane fuzzy sphere or giant graviton solution:
\begin{equation}
X^i_\mathrm{sphere} = \frac{\mu}{3} J^i ~,
\label{fuzzy}
\end{equation}
where $J^i$ satisfies the $SU(2)$ algebra,
\begin{equation}
[ J^i, J^j ] = i \epsilon^{ijk} J^k ~.
\label{su2}
\end{equation}
The reason why this solution is possible is basically because of the
fact that the matrix field $X^i$ feels an extra force due to the Myers
interaction which may stabilize the oscillatory force.  The fuzzy
sphere solution $X^i_\mathrm{sphere}$ preserves the full 16 dynamical
supersymmetries of the plane-wave and hence is 1/2-BPS object.  We
note that actually there is another fuzzy sphere solution of the form
$\frac{\mu}{6} J^i$.  However, it has been shown that such solution
does not have quantum stability and is thus non-BPS object
\cite{Sugiyama:2002bw}.

\section{Matrix model expansion around general background}
\label{bg-exp}

In this section, the plane-wave matrix model is expanded around the
general bosonic background, which is supposed to satisfy the classical
equations of motion, Eq. (\ref{eom}). 

We first split the matrix quantities into as follows:
\begin{equation}
X^I = B^I + Y^I ~,~~~ \Theta = F + \Psi ~,
\label{cl+qu}
\end{equation}
where $B^I$ and $F$ are the classical background fields while $Y^I$
and $\Psi$ are the quantum fluctuations around them.  The fermionic
background $F$ is taken to vanish from now on, since we will only
consider the purely bosonic background.  The quantum fluctuations are
the fields subject to the path integration.  In taking into account
the quantum fluctuations, we should recall that the matrix model
itself is a gauge theory.  This implies that the gauge fixing
condition should be specified before proceed further.  In this paper,
we take the background field gauge which is usually chosen in the
matrix model calculation as
\begin{equation}
D_\mu^{\rm bg} A^\mu_{\rm qu} \equiv
D_t A + i [ B^I, X^I ] = 0 ~.
\label{bg-gauge}
\end{equation}
Then the corresponding gauge-fixing $S_\mathrm{GF}$ and Faddeev-Popov
ghost $S_\mathrm{FP}$ terms are given by
\begin{equation}
S_\mathrm{GF} + S_\mathrm{FP} =  \int\!dt \,{\rm Tr}
  \left(
      -  \frac{1}{2} (D_\mu^{\rm bg} A^\mu_{\rm qu} )^2 
      -  \bar{C} \partial_{t} D_t C + [B^I, \bar{C}] [X^I,\,C]
\right) ~.
\label{gf-fp}
\end{equation}

Now by inserting the decomposition of the matrix fields (\ref{cl+qu})
into Eqs.~(\ref{pp-action}) and (\ref{gf-fp}), we get the gauge fixed
plane-wave action $S$ $(\equiv S_{pp} + S_\mathrm{GF} +
S_\mathrm{FP})$ expanded around the background.  The resulting acting
is read as
\begin{equation}
S =  S_0 + S_2 + S_3 + S_4 ~,
\end{equation}
where $S_n$ represents the action of order $n$ with respect to the
quantum fluctuations and, for each $n$, its expression is
\begin{align}
S_0 = \int dt \, \mathrm{Tr} \bigg[ \,
&      \frac{1}{2}(\dot{B}^I)^2  
        - \frac{1}{2} \left(\frac{\mu}{3}\right)^2 (B^i)^2 
        - \frac{1}{2} \left(\frac{\mu}{6}\right)^2 (B^a)^2 
        + \frac{1}{4}([B^I,\,B^J])^2
        - i \frac{\mu}{3} \epsilon^{ijk} B^i B^j B^k 
    \bigg] ~,
\notag \\
S_2 = \int dt \, \mathrm{Tr} \bigg[ \,
&       \frac{1}{2} ( \dot{Y}^I)^2 - 2i \dot{B}^I [A, \, Y^I] 
        + \frac{1}{2}([B^I , \, Y^J])^2 
        + [B^I , \, B^J] [Y^I , \, Y^J]
        - i \mu \epsilon^{ijk} B^i Y^j Y^k
\notag \\
&       - \frac{1}{2} \left( \frac{\mu}{3} \right)^2 (Y^i)^2 
        - \frac{1}{2} \left( \frac{\mu}{6} \right)^2 (Y^a)^2 
        + i \Psi^\dagger \dot{\Psi} 
        -  \Psi^\dagger \gamma^I [ \Psi , \, B^I ] 
        -i \frac{\mu}{4} \Psi^\dagger \gamma^{123} \Psi  
\notag \\ 
&       - \frac{1}{2} \dot{A}^2  - \frac{1}{2} ( [B^I , \, A])^2 
        + \dot{\bar{C}} \dot{C} 
        + [B^I , \, \bar{C} ] [ B^I ,\, C] \,
     \bigg] ~,
\notag \\
S_3 = \int dt \, \mathrm{Tr} \bigg[
&       - i\dot{Y}^I [ A , \, Y^I ] - [A , \, B^I] [ A, \, Y^I] 
        + [ B^I , \, Y^J] [Y^I , \, Y^J] 
        +  \Psi^\dagger [A , \, \Psi] 
\notag \\
&       -  \Psi^\dagger \gamma^I [ \Psi , \, Y^I ] 
        - i \frac{\mu}{3} \epsilon^{ijk} Y^i Y^j Y^k
        - i \dot{\bar{C}} [A , \, C] 
        +  [B^I,\, \bar{C} ] [Y^I,\,C]  \,
     \bigg] ~,
\notag \\
S_4 = \int dt \, \mathrm{Tr} \bigg[
&       - \frac{1}{2} ([A,\,Y^I])^2 + \frac{1}{4} ([Y^I,\,Y^J])^2 
     \bigg] ~.
\label{bgaction} 
\end{align}

Some comments are in order for the background gauge choice,
Eq. (\ref{bg-gauge}). One advantage of this gauge choice is that the
quadratic part of the action in terms of fluctuations, that is, the
quadratic action, is simplified.  In more detail, there appears the
term $-\frac{1}{2} ([ B^I, Y^I])^2$ in the expansion of the potential
$\frac{1}{4} ([X^I, X^J])^2$, which is canceled exactly by the same
term with the opposite sign coming from the gauge fixing term of
Eq. (\ref{gf-fp}) and hence absent in $S_2$ of Eq. (\ref{bgaction}).
This cancellation has given some benefits in the actual flat space
matrix model calculation.  This is also the case in the present
plane-wave matrix model, except however for the potential of $Y^i$.
As we will see later, $-\frac{1}{2} ([ B^i, Y^i])^2$ is responsible
for completing the $Y^i$ potential into the nice square form.  Thus
the same term with the opposite sign from the gauge fixing term
remains in the quadratic action.  At later stage, the presence of this
term will have an important implication in taking into account of the
unphysical gauge degrees of freedom which are eventually eliminated by
those of ghosts.

\section{Fuzzy sphere configuration and fluctuations}
\label{fuzzy-config}

We now set up the background configuration for the membrane fuzzy
spheres.  Since we will study the interaction of two fuzzy spheres,
the matrices representing the background have the $2 \times 2$ block
diagonal form as
\begin{equation}
B^I = \begin{pmatrix} B_{(1)}^I & 0 \\ 0 & B_{(2)}^I
      \end{pmatrix} ~,
\label{gb}
\end{equation}
where $B_{(s)}^I$ with $s=1,2$ are $N_s \times N_s$ matrices.  If we
take $B^I$ as $N \times N$ matrices, then $N = N_1 + N_2$.  

\begin{figure}
\begin{center}
 \includegraphics[scale=.8]{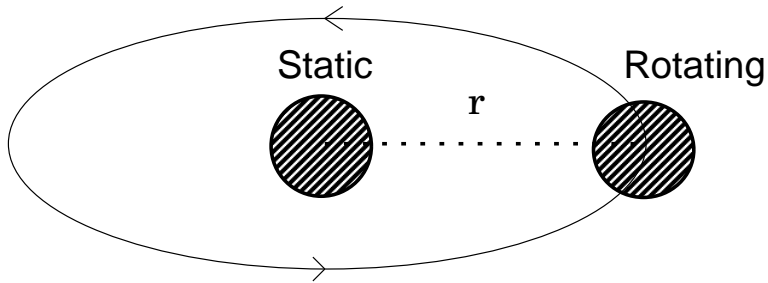}
\end{center}
\caption{Schematic view of the configuration for two membrane fuzzy
spheres.  The plane of the circular motion is $x^4$-$x^5$ plane, and
the fuzzy spheres are actually points in this plane.}
\label{config}
\end{figure}

The two fuzzy spheres are taken to be static in the space where they
span, and hence represented by the classical solution, (\ref{fuzzy});
\begin{equation}
B_{(s)}^i = \frac{\mu}{3} J_{(s)}^i ~,
\label{b3}
\end{equation}
where, for each $s$, $J_{(s)}^i$ is in the $N_s$-dimensional
irreducible representation of $SU(2)$ and satisfies the $SU(2)$
algebra, Eq. (\ref{su2}).  In the $SO(6)$ symmetric transverse space,
the fuzzy spheres are regarded as point objects, of course, in a sense
of ignoring the matrix nature.  We first let the second fuzzy sphere
given by the background of $s=2$ be at the origin in the transverse
space and stay there.  As for the first fuzzy sphere, it is made to
move around the second sphere in the form of circular motion with the
radius $r$.  Obviously, this configuration is one of the classical
solutions of the equations of motion as one can see from
Eq.~(\ref{osc}).  Recalling that the transverse space is $SO(6)$
symmetric, all the possible choices of two-dimensional sub-plane where
the circular motion takes place are equivalent.  Thus, without loss of
generality, we can take a certain plane for the circular motion.  In
this paper, the $x^4$-$x^5$ plane is chosen.  Then the configuration
in the transverse space is given by
\begin{equation}
B^4_{(1)} = r \cos \left( \frac{\mu}{6} t \right) 
             {\bf 1}_{N_1 \times N_1}~,~~~
B^5_{(1)} = r \sin \left( \frac{\mu}{6} t \right) 
             {\bf 1}_{N_1 \times N_1}~.
\label{b6}
\end{equation}
Eqs.~(\ref{b3}) and (\ref{b6}) compose the background configuration
about which we are concerned, and all other elements of matrices $B^I$
are set to zero.  We would like to note that not only the fuzzy sphere
at the origin given by $B^I_{(2)}$ but also the rotating one,
$B^I_{(1)}$, is supersymmetric \cite{Park:2002cb}.  A schematic view
of the background configuration is presented in Fig.~\ref{config}.

If we evaluate the classical value of the action for this background,
it is zero;
\begin{equation}
S_0 = 0 ~.
\label{clav}
\end{equation}
From now on, we are going to compute the one-loop correction to this
action, that is, to the background, (\ref{b3}) and (\ref{b6}), due to
the quantum fluctuations via the path integration of the quadratic
action $S_2$, and obtain the one-loop effective action
$\Gamma_\mathrm{eff}$ or the effective potential $V_\mathrm{eff}$ as a
function of $r$, the radius of the circular motion.

For the justification of one-loop computation or the semi-classical
analysis, it should be made clear that $S_3$ and $S_4$ of
Eq.~(\ref{bgaction}) can be regarded as perturbations.  For this
purpose, following \cite{Dasgupta:2002hx}, we rescale the fluctuations
and parameters as
\begin{gather}
A   \rightarrow \mu^{-1/2} A   ~,~~~
Y^I \rightarrow \mu^{-1/2} Y^I ~,~~~
C   \rightarrow \mu^{-1/2} C   ~,~~~
\bar{C} \rightarrow \mu^{-1/2} \bar{C} ~,
\notag \\
r \rightarrow \mu r ~,~~~ t \rightarrow \mu^{-1} t ~.
\label{rescale}
\end{gather}
Under this rescaling, the action $S$ in the background (\ref{b3}) and
(\ref{b6}) becomes
\begin{align}
S =  S_2 + \mu^{-3/2} S_3 + \mu^{-3} S_4 ~,
\label{ssss}
\end{align}
where $S_2$, $S_3$ and $S_4$ do not have $\mu$ dependence.  Now it is
obvious that, in the large $\mu$ limit, $S_3$ and $S_4$ can be treated
as perturbations and the one-loop computation gives the sensible
result.

Based on the structure of (\ref{gb}), we now write the quantum
fluctuations in the $2 \times 2$ block matrix form as follows.
\begin{gather}
A   = \begin{pmatrix}
          Z_{(1)}^0           &   \Phi^0      \\
         \Phi^{0 \dagger} &   Z_{(2)}^0
      \end{pmatrix} ~,~~~
Y^I = \begin{pmatrix}
         Z_{(1)}^I           &   \Phi^I      \\
         \Phi^{I \dagger} &   Z_{(2)}^I
      \end{pmatrix} ~,~~~
\Psi = \begin{pmatrix}
         \Psi_{(1)} &  \chi \\
         \chi^{\dagger} & \Psi_{(2)} 
       \end{pmatrix} ~,
 \notag \\
C =    \begin{pmatrix}
         C_{(1)}  &  C  \\ 
         C^{\dagger} & C_{(2)} 
       \end{pmatrix} ~,~~~
\bar{C} = \begin{pmatrix}
              \bar{C}_{(1)} &  \bar{C}  \\
              \bar{C}^\dagger &  \bar{C}_{(2)}
           \end{pmatrix} ~.
\label{q-fluct}
\end{gather}
Although we denote the block off-diagonal matrices for the ghosts by
the same symbols with those of the original ghost matrices, there will
be no confusion since $N \times N$ matrices will never appear in what
follows.  The above form of matrices is convenient, since the block
diagonal and block off-diagonal parts decouple from each other in the
quadratic action and thus can be taken into account separately at
one-loop level;
\begin{equation}
S_2 = S_\text{diag} + S_\text{off-diag} ~,
\label{s2}
\end{equation}
where $S_\text{diag}$ ($S_\text{off-diag}$) implies the action for the
block (off-) diagonal fluctuations.  We note that there is no $\mu$
parameter in $S_\text{diag}$ and $S_\text{off-diag}$ due to the above
rescaling (\ref{rescale}).

\section{Stability of fuzzy sphere}
\label{stable}

In this section, the path integration of $S_\mathrm{diag}$ is
performed.  The resulting effective action will enable us to check the
one-loop stability of the fuzzy sphere configuration, Eqs.~(\ref{b3})
and (\ref{b6}).  In fact, the background encoding the circular motion,
Eq.~(\ref{b6}), does not contribute to $S_\mathrm{diag}$ basically
because of the fact that it is proportional to the identity matrix.
This leads to the situation that the quantum fluctuations around one
fuzzy sphere do not interact with those around another fuzzy sphere,
and the effective action is just the sum of that for each fuzzy
sphere.  This can then be stated as
\begin{equation}
S_\mathrm{diag} = 
\int dt \sum^2_{s=1} ( L^B_{(s)} + L^F_{(s)} + L^G_{(s)} ) ~,
\label{diag-a}
\end{equation}
where three Lagrangians, $L^B_{(s)}$, $L^F_{(s)}$, and $L^G_{(s)}$,
are those for the bosonic, fermionic, and ghost fluctuations
respectively around the $s$-th fuzzy sphere.  We note that, at
quadratic level, there is no mixing between these three kinds of
fluctuations.

\subsection{Bosonic fluctuation}

We first evaluate the path integral of bosonic fluctuations.
The corresponding Lagrangian is given by
\begin{align}
L^B_{(s)} = \frac{1}{2} \mathrm{Tr} \bigg[
& - \left( \dot{Z}_{(s)}^0 \right)^2 
  - \frac{1}{3^2}
       \left( \big[ J_{(s)}^i, Z_{(s)}^0 \big] \right)^2
  + \left( \dot{Z}_{(s)}^i \right)^2 
  - \frac{1}{3^2}
       \left( 
            Z_{(s)}^i 
          + i \epsilon^{ijk} \big[ J_{(s)}^j, Z_{(s)}^k \big] 
       \right)^2
                       \notag \\
& + \frac{1}{3^2}
       \left( \big[ J_{(s)}^i, Z_{(s)}^i \big] \right)^2
  + \left( \dot{Z}_{(s)}^a \right)^2 
  - \frac{1}{3^2}
    \left( 
          \frac{1}{4} \left( Z_{(s)}^a \right)^2 
          - \left( \big[ J_{(s)}^i, Z_{(s)}^a \big] \right)^2
    \right) \bigg] ~,
\label{diag-lb}
\end{align}
where, as alluded to at the end of section \ref{bg-exp}, the quadratic
term of $[ J_{(s)}^i, Z_{(s)}^i ]$ coming from the gauge fixing term
appears because the same term with opposite sign has been used for
making the complete square form $( Z_{(s)}^i + i \epsilon^{ijk} [
J_{(s)}^j, Z_{(s)}^k ] )^2$.

In order to perform the actual path integration, it is useful to
diagonalize the fluctuation matrices and obtain the mass spectrum.  By
the way, since the diagonalization itself has been already given in
\cite{Dasgupta:2002hx}, we will be brief in its presentation and
present only the essential points.

The starting point is the observation that the mass terms are written
in terms of the commutators with $J_{(s)}^i$ satisfying
Eq.~(\ref{su2}), the $SU(2)$ algebra.  This indicates that we can use
the representation theory of $SU(2)$ for the diagonalization.  We
regard an $N_s \times N_s$ matrix as an $N_s^2$-dimensional reducible
representation of $SU(2)$, which decomposes into irreducible spin $j$
representations with the range of $j$ from $0$ to $N_s -1$, that is,
$N_s^2 = 1 \oplus 3 \oplus \cdots \oplus (2 N_s -1 )$.  Based on this
decomposition, an $N_s \times N_s$ matrix may be expanded as
\begin{equation}
Z_{(s)} = \sum^{N_s-1}_{j=0} \sum^{j}_{m=-j} z_{(s) jm} Y_{jm}^{(s)} ~,
\label{mode}
\end{equation}
where the $N_s \times N_s$ matrix $Y_{jm}^{(s)}$ is the matrix
spherical harmonics transforming in the irreducible spin $j$
representation and $z_{(s)jm}$ is the corresponding spherical mode.
We note that $z_{(s)jm}$ satisfies the following reality condition
since the fluctuation matrices are Hermitian.
\begin{equation}
z_{(s)jm}^* = (-1)^m z_{(s)j \, -m} ~.
\label{real}
\end{equation}
The expansion of matrices in the $SU(2)$ language enables us to use
the properties of the $SU(2)$ generators, which are given by
\begin{align}
&  [ J_{(s)}^3, \, Y_{jm}^{(s)} ] = m Y_{jm}^{(s)} ~,
&& [ J_{(s)}^+, \, Y_{jm}^{(s)} ] 
      = \sqrt{ (j-m)(j+1+m) } \; Y_{j \, m+1}^{(s)} ~,
   \notag \\
&  [J_{(s)}^i, \, [J_{(s)}^i , \, Y_{jm}^{(s)} ]] 
      = j(j+1) Y_{jm}^{(s)} ~,
&& [ J_{(s)}^-, \, Y_{jm}^{(s)} ] 
      = \sqrt{ (j+m)(j+1-m) } \; Y_{j \, m-1}^{(s)} ~,
\label{alg}
\end{align}
where $J_{(s)}^\pm = J_{(s)}^1 \pm i J_{(s)}^2$.  For the
normalization of the matrix spherical harmonics, we choose
\begin{equation}
\mathrm{Tr} ( Y^{(s) \dagger}_{j'm'} Y_{jm}^{(s)}) 
= N_s \delta_{j'j} \delta_{m'm} ~.
\label{norm}
\end{equation}
Having equipped with the necessary machinery, we now proceed the
diagonalization.  

By direct application of Eqs.~(\ref{mode}), (\ref{alg}), and
(\ref{norm}), the gauge field fluctuation $Z^0_{(s)}$ and the
fluctuations in the $SO(6)$ directions $Z^a_{(s)}$ are immediately
diagonalized.  The corresponding spherical modes and their masses are
\begin{align}
& z^0_{(s)jm} && \frac{1}{3} \sqrt{j(j+1)}  ~,
\notag \\
& z^a_{(s)jm} && \frac{1}{3} (j+ \frac{1}{2}) ~,
\label{d-so6}
\end{align}
where $0 \le j \le N_s-1$ and $-j \le m \le j$.

As for the fluctuations in the $SO(3)$ directions, we should first
consider the potential $\frac{1}{2 \cdot 3^2} ( Z_{(s)}^i + i
\epsilon^{ijk} [ J_{(s)}^j, Z_{(s)}^k ] )^2$ and diagonalize it by
solving the eigenvalue problem,
\begin{equation}
Z_{(s)}^i + i \epsilon^{ijk} [ J_{(s)}^j, Z_{(s)}^k ] 
= \lambda Z_{(s)}^i ~.
\end{equation}
By defining the combinations of matrices $Z_{(s)}^\pm = Z_{(s)}^1 \pm
i Z_{(s)}^2$ and using Eqs.~(\ref{mode}) and (\ref{alg}), it turns out
that the eigenvalue $\lambda$ takes the values of $-j$, $j+1$, or $0$.
Let us now denote the spherical eigenmodes of matrix eigenvectors as
$u_{(s)jm}$, $v_{(s)jm}$, and $w_{(s)jm}$ for $\lambda = -j$, $j+1$,
and $0$, respectively.  The eigenmodes satisfy the reality condition
in the form of Eq.~(\ref{real}).  Then the fluctuation matrices may be
expressed with respect to these eigenmodes for each eigenvalue.

For $\lambda = -j$, we have the following expressions:
\begin{align}
Z^+_{(s)} 
& = - \frac{1}{\sqrt{N_s}}
    \sqrt{ \frac{(j+m) (j+1+m)}{j(2j+1)} } \;
    u_{(s) j-1 \, m} Y_{j \, m+1}^{(s)} ~,
  \notag \\
Z^-_{(s)}
&= \frac{1}{\sqrt{N_s}}
    \sqrt{ \frac{(j-m) (j+1-m)}{j(2j+1)} } \;
    u_{(s) j-1 \, m} Y_{j \, m-1}^{(s)} ~,
  \notag \\
Z^3_{(s)}
&= \frac{1}{\sqrt{N_s}}
    \sqrt{ \frac{(j+m) (j-m)}{j(2j+1)} } \;
    u_{(s) j-1 \, m} Y_{j m}^{(s)} ~,
\label{u}
\end{align}
where $0 < j < N_s$, $-j < m < j$, and the normalization constant are
chosen so that the kinetic term for the mode $u_{(s)jm}$ is of the
form $\frac{1}{2} (\dot{u}_{(s)jm})^2$.  Here and in what follows, the
summations over $j$ and $m$ with specified ranges are implicit.

For $\lambda = j+1$, we have
\begin{align}
Z^+_{(s)} 
& = \frac{1}{\sqrt{N_s}}
    \sqrt{ \frac{(j-m) (j+1-m)}{(j+1)(2j+1)} } \;
    v_{(s) j+1 \, m} Y_{j \, m+1}^{(s)} ~,
  \notag \\
Z^-_{(s)}
&= - \frac{1}{\sqrt{N_s}}
    \sqrt{ \frac{(j+m) (j+1+m)}{(j+1)(2j+1)} } \;
    v_{(s) j+1 \, m} Y_{j \, m-1}^{(s)} ~,
  \notag \\
Z^3_{(s)}
&= \frac{1}{\sqrt{N_s}}
    \sqrt{ \frac{(j+1+m) (j+1-m)}{(j+1)(2j+1)} } \;
    v_{(s) j+1 \, m} Y_{j m}^{(s)} ~,
\label{v}
\end{align}
where $0 \le j < N_s$, $-j-1 \le m \le j+1$, and the normalization
constant are chosen in the same way with the case of $u_{(s)jm}$.

Finally, for $\lambda = 0$, 
\begin{equation}
Z_{(s)}^i = \frac{1}{\sqrt{N_s}} \frac{ w_{(s)jm} }{ \sqrt{j(j+1)} }
     [ J_{(s)}^i, Y_{jm}^{(s)} ]
\label{w}
\end{equation} 
with $0 < j < N_s$ and $-j \le m \le j$.  As pointed out in
\cite{Dasgupta:2002hx}, the modes for $\lambda = 0$ case correspond to
the degrees of freedom for the gauge transformation and are thus
unphysical.  Since the authors of \cite{Dasgupta:2002hx} took the
physical Weyl gauge, $A=0$, and worked in the operator formulation, it
was not necessary to consider these unphysical modes seriously.
However, they should be involved properly in the present context
because our gauge choice is the covariant background gauge and we work
in the path integral formulation.

We turn to the remaining potential term in the Lagrangian
(\ref{diag-lb}) which is the square of $[ J_{(s)}^i, Z_{(s)}^i ]$.
Interestingly enough, the matrices expanded in terms of the eigenmodes
$u_{(s)jm}$ and $v_{(s)jm}$, Eqs.~(\ref{u}) and (\ref{v}), do not give
any contribution, since we obtain
\begin{equation}
[ J_{(s)}^i, Z_{(s)}^i ] = 0 \, ,
\end{equation}
for $\lambda = -j$ and $j+1$.  Only the modes corresponding to
$\lambda = 0$ contribute to the potential, which is evaluated as
\begin{equation}
\frac{1}{2 \cdot 3^2} j (j+1) | w_{(s)jm} |^2 ~,
\end{equation}
for each $j$ and $m$.

Having diagonalized the fluctuations $Z^i_{(s)}$, we get the following
list of spherical modes in the $SO(3)$ directions with their masses
and the ranges of spin $j$:
\begin{align}
&u_{(s)jm} && \frac{1}{3} (j+1)         && 0 \le j \le N_s - 2 ~,
\notag \\
&v_{(s)jm} && \frac{1}{3} j             && 1 \le j \le N_s  ~,
\notag \\
&w_{(s)jm} && \frac{1}{3} \sqrt{j(j+1)} && 1 \le j \le N_s -1 ~,
\label{d-so3}
\end{align}
where $-j \le m \le j$.  

With respect to the spherical modes of Eqs.~(\ref{d-so6}) and
(\ref{d-so3}), the Lagrangian (\ref{diag-lb}) is then written in the
diagonalized form as
\begin{align}
L^B_{(s)} =
& \frac{1}{2}\sum_{j = 0}^{N_s -1} \left( - | \dot{z}^0_{(s) jm} |^2 
  + \frac{1}{3^2} j (j+1) | z^0_{(s) jm} |^2
  + | \dot{z}^a_{(s) jm} |^2 
  - \frac{1}{3^2} \left( j+\frac{1}{2} \right)^2
    | z^a_{(s) jm} |^2 \right)
   \notag \\
& + \frac{1}{2} \sum_{j=0}^{N_s - 2} \left( | \dot{u}_{(s) jm} |^2
  - \frac{1}{3^2} ( j+ 1 )^2 | u_{(s) jm} |^2
    \right)
  + \frac{1}{2}\sum_{j=1}^{N_s} \left( | \dot{v}_{(s) jm} |^2
  - \frac{1}{3^2} j^2 | v_{(s) jm} |^2 \right)
   \notag \\
& + \frac{1}{2}\sum_{j = 1}^{N_s -1} \left( - | \dot{w}_{(s) jm} |^2 
  + \frac{1}{3^2} j (j+1) | w_{(s) jm} |^2
    \right) ~,
\end{align}
where the sum over $m$ with the range $ -j \le m \le j$ is understood.
The Lagrangian is just the sum of various harmonic oscillator
Lagrangians, which are non-interacting with each other, and therefore
the path integration is now straightforward.  As a result, what we
obtain is
\begin{align}
 & \prod_{j=0}^{N_s-1} 
  \left[
    \det \left( \partial_t^2 + \frac{1}{3^2} j(j+1)
        \right)
   {\det}^6 \bigg( \partial_t^2 + \frac{1}{3^2}
                   \left( j+\frac{1}{2} \right)^2
            \bigg)
  \right]^{-j - \frac{1}{2}}
\notag \\
\times & \prod_{j'=0}^{N_s-2} 
        \left[ 
              \det \left( \partial_t^2 
                         + \frac{1}{3^2} ( j'+ 1 )^2
                   \right)
        \right]^{-j'-\frac{1}{2}} 
        \prod_{j''=1}^{N_s} 
        \left[ 
              \det \left( \partial_t^2 
                          + \frac{1}{3^2} j''^2
                   \right)
        \right]^{-j''-\frac{1}{2}}
\notag \\
\times & \prod_{j'''=1}^{N_s-1} 
        \left[
              \det \left( \partial_t^2 
                       + \frac{1}{3^2} j''' (j'''+1)
                   \right)
        \right]^{-j'''-\frac{1}{2}} ~.
\label{p-db}
\end{align}

\subsection{Fermionic fluctuation}

We turn to the path integral of fermionic fluctuations.  The
Lagrangian is written as
\begin{equation}
L^F_{(s)} = \mathrm{Tr} 
\left(
   i\Psi_{(s)}^{\dagger}\dot{\Psi}_{(s)} 
   - \frac{1}{3} \Psi_{(s)}^{\dagger} 
                   \gamma^i [\Psi_{(s)},J_{(s)}^i] 
 - i \frac{1}{4} \Psi_{(s)}^{\dagger}\gamma^{123}\Psi_{(s)}
\right) \, ,
\end{equation}
It is convenient for our calculation of fermionic part to introduce
the $SU(2) \times SU(4)$ formulation since the preserved symmetry in
the plane-wave matrix model is $SO(3)\times SO(6) \sim SU(2)\times
SU(4)$ rather than $SO(9)$.  In this formulation the $SO(9)$ spinor
$\Psi_{(s)}$ is decomposed as
\begin{equation}
\begin{matrix}
{\bf 16}   & \rightarrow & ({\bf 2}, {\bf 4})  &+ & 
                         (\bar{\bf 2}, \bar{\bf 4})  \\
\Psi_{(s)} & \rightarrow & \psi_{(s)A\alpha}   &  &
                          \psi^{\dagger B \beta}_{(s)} \,, 
\label{s-dec}
\end{matrix}
\end{equation}
where $A$ implies a fundamental $SU(4)$ index and $\alpha$ is a
fundamental $SU(2)$ index.  According to this decomposition, we may
take the expression of $\Psi_{(s)}$ as
\begin{equation}
\Psi_{(s)} = \frac{1}{\sqrt{2}}
             \begin{pmatrix}
                 \psi_{(s)A\alpha}     \\
                 \epsilon_{\alpha\beta}
                    \psi_{(s)}^{\dagger A\beta}
             \end{pmatrix} \, .
\label{d-dec}
\end{equation}
We also rewrite the $SO(9)$ gamma matrices $\gamma^I$'s 
in terms of $SU(2)$ and $SU(4)$ ones as follows:  
\begin{equation}
\gamma^i = \begin{pmatrix}
               - \sigma^i \times 1 &         0        \\
                         0         & \sigma^i \times 1 
           \end{pmatrix} \,, \quad 
\gamma^a = \begin{pmatrix}
                         0                 & 1\times \rho^a  \\
              1 \times (\rho^a)^{\dagger}  &       0
           \end{pmatrix} \, ,
\end{equation}
where the $\sigma^i$'s are the standard $2 \times 2$ Pauli matrices
and six of $\rho^a$ are taken to form a basis of $4 \times 4$
anti-symmetric matrices. The original $SO(9)$ Clifford algebra is
satisfied as long as we take normalizations so that the gamma matrices
$\rho^a$ with $SU(4)$ indices satisfy the algebra
\begin{align}
\rho^a(\rho^b)^{\dagger} + \rho^b(\rho^a)^{\dagger} = 2\delta^{ab}\,.
\label{cliff}
\end{align} 

By introducing the above $SU(2) \times SU(4)$ formulation, 
the Lagrangian for the fermionic fluctuations is rewritten as
\begin{align}
L^F_{(s)} = \mathrm{Tr} 
\left(
     i\psi_{(s)}^{\dagger A\alpha}\dot{\psi}_{(s) A\alpha} 
   + \frac{1}{3}\psi^{\dagger A\alpha}_{(s)} 
            (\sigma^i)_{\alpha}^{~\beta}
            [\psi_{(s) A\beta}, J^i_{(s)}] 
   - \frac{1}{4}\psi_{(s)}^{\dagger A\alpha} \psi_{(s) A\alpha}
\right) \, .
\label{diag-lf}
\end{align} 

The diagonalization of the Lagrangian proceeds in the same way as 
in the previous subsection.  In the present case, it is achieved
by solving the following eigenvalue problem:
\begin{align}
(\sigma^i)_{\alpha}{}^\beta [ J^i_{(s)}, \psi_{(s) A \beta} ] 
= \lambda \psi_{(s) A \alpha} \, .
\end{align}
We first expand the $N_s \times N_s$ fermionic matrix
$\psi_{(s)A\alpha}$ in terms of the matrix spherical harmonics as
\begin{align}
\psi_{(s)A\alpha} = \sum_{j=0}^{N_s-1} \sum_{m=-j}^{j}
          \psi_{(s)A\alpha}^{jm}Y_{jm}^{(s)} \,.
\end{align}
If we plug this expansion into the above eigenvalue equation and use
the $SU(2)$ algebra Eq.~(\ref{alg}), we see that the eigenvalues are
$\lambda = j$ and $\lambda = -j-1$ \cite{Dasgupta:2002hx}.  Let us now
introduce the fermionic spherical eigenmodes $\eta_{(s)}^{jm}$ and
$\pi_{(s)}^{jm}$ corresponding to $\lambda = j$ and $-j-1$
respectively.  As for the spinorial structure, $\eta_{(s)}^{jm}$
($\pi_{(s)}^{jm}$) carries the (anti) fundamental $SU(4)$ index.  We
note that, from now on, we suppress the $SU(4)$ indices.

Then, as the eigenstate for $\lambda = j$, the matrix
$\psi_{(s)\alpha}$ has the expansion in terms of the eigenmode
$\eta^{jm}$ as
\begin{align}
\psi_{(s)+} =& \frac{1}{\sqrt{N_s}}
               \sqrt{\frac{j+1+m}{2j+1}} \,
               \eta_{(s)}^{j + \frac{1}{2}~m + \frac{1}{2}} \,
               Y_{jm}^{(s)} \,, 
\notag \\
\psi_{(s)-} =& \frac{1}{\sqrt{N_s}}
               \sqrt{\frac{j-m}{2j+1}} \,
               \eta_{(s)}^{j + \frac{1}{2}~m + \frac{1}{2}} \,
               Y_{j \, m+1}^{(s)} \,,
\label{df-1}
\end{align}
where $0 \le j \le N_s-1$, $-j-1 \le m \le j$, and the subscripts
$\pm$ denote the $SU(2)$ indices measured by $\sigma^3$.

On the other hand, for $\lambda = -j-1$, we have
\begin{align}
\psi_{(s)+} =& - \frac{1}{\sqrt{N_s}}
               \sqrt{\frac{j-m}{2j+1}} \,
               (\pi_{(s)}^{\dagger})^{ j-\frac{1}{2}\, 
                                       m + \frac{1}{2} } \,
               Y_{jm}^{(s)} \,, 
\notag \\
\psi_{(s)-} =& \frac{1}{\sqrt{N_s}}
               \sqrt{\frac{j+1+m}{2j+1}} \,
               (\pi_{(s)}^{\dagger})^{j-\frac{1}{2} \, 
                                      m + \frac{1}{2} } \,
               Y_{j \, m+1}^{(s)}\,,
\label{df-2}
\end{align}
where $1 \le j \le N_s-1$ and $-j \le m \le j-1$. 


By using the mode-expansions, Eqs.~(\ref{df-1}) and (\ref{df-2}), and
the $SU(2)$ algebra (\ref{alg}), the fermionic Lagrangian
(\ref{diag-lf}) becomes
\begin{align}
L^F_{(s)} =
& \sum_{j=\frac{1}{2}}^{N_s-\frac{3}{2}} 
 \left(
    i\pi_{(s)jm}^\dagger \dot{\pi}_{(s) jm} 
    - \frac{1}{3} \left(j + \frac{3}{4}\right) 
         \pi^\dagger_{(s)jm} \pi_{(s)jm}
 \right) 
\notag \\
& + \sum_{j=\frac{1}{2}}^{N_s-\frac{1}{2}}
 \left(
    i\eta^\dagger_{(s)jm} \dot{\eta}_{(s)jm} 
    - \frac{1}{3} \left(j + \frac{1}{4}\right)
          \eta_{(s)jm}^\dagger \eta_{(s)jm}
 \right) \, ,
\end{align}
where it should be understood that there is the summation over $m$
with the range $-j \le m \le j$.  Now, the path integration for this
Lagrangian may be evaluated immediately, and gives
\begin{align}
& \prod_{j=\frac{1}{2}}^{N_{s} - \frac{3}{2}}
  \left[
      \det \left( \partial_t^2 
                 + \frac{1}{3^2}
                   \left(j +\frac{3}{4}\right)^2
           \right)
  \right]^{2 (2j+1)} 
\notag \\
\times &
  \prod_{j' =\frac{1}{2}}^{N_{s} - \frac{1}{2}}
  \left[
      \det \left( \partial_t^2 
                 + \frac{1}{3^2}
                   \left(j' +\frac{1}{4}\right)^2
           \right)
  \right]^{2 (2j'+1)} \,. 
\label{p-df}
\end{align}

\subsection{Ghost fluctuation}

As the final part of the diagonal fluctuations, we consider the
Lagrangian for the ghost fluctuations, which is given by
\begin{align}
L^G_{(s)} = \mathrm{Tr}
\left(  \dot{\bar{C}}_{(s)}\dot{C}_{(s)} 
      + \frac{1}{3^2}
        [J^i_{(s)}, \bar{C}_{(s)}][J^i_{(s)}, C_{(s)}]
\right) ~.
\end{align}

By using the $SU(2)$ algebra (\ref{alg}) and the following expansions
in terms of the matrix spherical harmonics
\begin{align}
C_{(s)} = \frac{1}{\sqrt{N_{(s)}}}c_{(s)jm} Y_{jm}^{(s)} ~,~~~
\bar{C}_{(s)} 
        = \frac{1}{\sqrt{N_{(s)}}} \bar{c}_{(s)jm}Y_{jm}^{(s)} \,,
\end{align}
with $0 \le j \le N_s-1$ and $-j \le m \le j$,
we may rewrite the above Lagrangian as 
\begin{align}
L^G_{(s)} = \sum_{j=0}^{N_s-1} 
 \left( \dot{\bar{c}}^*_{(s)jm} \dot{c}_{(s)jm}
        - \frac{1}{3^2} j(j+1) 
          \bar{c}^*_{(s)jm} c_{(s)jm}
 \right) \, ,
\end{align}
where the sum over $m$ is implicit.  Then the result of the path
integration for this Lagrangian is
\begin{equation}
\prod_{j=0}^{N_s-1} 
\left[
    \det \left(  \partial_t^2 + \frac{1}{3^2} j(j+1) \right)
\right]^{2j+1} ~.
\label{p-dg}
\end{equation}

\subsection{One-loop stability}

In the previous subsections, we have evaluated the path integrals for
the block diagonal fluctuations of the action $S_\text{diag}$,
(\ref{diag-a}).  Thus, we may now consider the effective action.  As
can be inferred from Eq.~(\ref{diag-a}), it is enough to consider only
the effective action of a given fuzzy sphere.  If we let
$\Gamma^{(s)}_\text{eff}$ be the effective action for the $s$-th fuzzy
sphere, then it is given by
\begin{align}
e^{i\Gamma^{(s)}_\text{eff}} 
= (\ref{p-db}) \times (\ref{p-df}) \times (\ref{p-dg}) ~.
\label{e-d-eff}
\end{align}
The right hand side may be viewed just as the product of determinants
of non-interacting quantum mechanical simple harmonic oscillators with
various frequencies.  In fact, as we will see in the next section,
this is also the case for the effective action describing the
interaction between two fuzzy spheres.  Thus, it is worthwhile to
consider a generic situation, which is useful both in the present and
the next section.

Let us then consider a situation,
\begin{equation}
e^{i \Gamma_\mathrm{eff}} 
= \prod_n {\det}^{a_n} ( \partial_t^2 - 2 i p_n \partial_t + m_n^2 ) ~,
\label{gen}
\end{equation}
where $p_n$ is included for the later usage.  The formal expression of
the effective action is read as
\begin{equation}
\Gamma_\mathrm{eff} = -i \sum_n a_n \ln 
            \det ( \partial_t^2 - 2 i p_n \partial_t + m_n^2 ) ~.
\label{g-eff}
\end{equation}
By using the relation $\ln \det M = \mathrm{Tr} \ln M$ for a given
matrix $M$, we may present a prototype calculation of single
determinant as
\begin{align}
\mathrm{Tr} \ln ( \partial_t^2 - 2 i p \partial_t + m^2 )
& = \int dt \langle t | \ln 
        ( \partial_t^2 - 2 i p \partial_t + m^2 ) | t \rangle
  \notag \\
& = \int dt \int^\infty_{-\infty} \frac{dk}{2 \pi} 
         \ln ( -k^2 + 2 p k + m^2 )
  \notag \\
& = i \int dt \sqrt{m^2 + p^2} ~,
\end{align}
where the momentum integration has appeared by inserting the momentum
space identity, $\int dk | k \rangle \langle k | =1$, between bra and
ket vectors for time, and the final result has been derived by using
the formula
\begin{equation}
\int^\infty_{-\infty} \frac{dk}{2 \pi} 
\ln ( -k^2 + 2 pk + m^2 - i \epsilon) = i \sqrt{ m^2 + p^2} ~.
\end{equation}
If we apply the result of this single determinant calculation to
Eq.~(\ref{g-eff}), then the effective action is finally obtained as
\begin{align}
\Gamma_\mathrm{eff} = \int dt  \sum_n a_n \sqrt{ m_n^2 + p_n^2} ~.
\label{gen-eff}
\end{align}

We now return to the present case.  First of all, we observe that the
ghost part (\ref{p-dg}) eliminates two determinant factors of the
bosonic result (\ref{p-db}) which are actually the contributions from
the unphysical modes.  Thus, what we get from the bosonic and the
ghost parts is only the determinant factors from the physical bosonic
modes.  By using the generic expression, Eq.~(\ref{gen-eff}), it turns
out that their contributions to the effective action are
\begin{equation}
- \frac{1}{3^2} N_s ( 8 N_s^2+1 ) ~.
\end{equation}
On the other hand, the contribution from the fermionic part
(\ref{p-df}) is obtained as
\begin{equation}
+ \frac{1}{3^2} N_s ( 8 N_s^2+1 ) ~.
\end{equation}
Therefore, there is no net contribution to the effective action and we
can conclude that the one loop effective action obtained after
integrating out each of the block diagonal fluctuations vanishes;
\begin{equation}
\Gamma^{(s)}_\mathrm{eff} = 0 ~.
\end{equation}
This indicates that each membrane fuzzy sphere has quantum stability
at least at one-loop level, by which we mean that the fuzzy sphere
does not receive quantum corrections.  We note that, if we take $N_s
=2$ in the above two contributions, the result of the previous path
integral computation \cite{Sugiyama:2002bw} is recovered.

\section{Interaction between fuzzy spheres}
\label{interact}

We turn to the action $S_\text{off-diag}$ in Eq.~(\ref{s2}) for the
block off-diagonal fluctuations and compute the one-loop effective
potential describing the interaction between two fuzzy spheres.  The
action is given by
\begin{align}
S_\text{off-diag} = \int dt ( L^B + L^F + L^G ) ~,
\label{oda}
\end{align}
where $L^B$, $L^F$ and $L^G$ are the Lagrangians for the off-diagonal
bosonic, fermionic and ghost fluctuations of Eq.~(\ref{q-fluct})
respectively and their explicit expressions will be presented in due
course.  Since, as in the previous section, there is no mixing between
different kinds of fluctuations, each Lagrangian can be considered
independently.  

In calculating the effective action, the prescription given by Kabat
and Taylor \cite{Kabat:1998im} is usually used.  We note however that,
at the present situation, it is more helpful to use the expansion in
terms of the matrix spherical harmonics as in the previous section.

\subsection{Bosonic fluctuation}

Let us first consider the bosonic Lagrangian and evaluate its path
integral.  The Lagrangian is
\begin{align}
L^B = \mathrm{Tr}  
\bigg\{ 
& - | \dot{\Phi}^0 |^2 
  + r^2 | \Phi^0 |^2
  + \frac{1}{3^2} \Phi^{0 \dagger} J^i \circ ( J^i \circ \Phi^0 )
 \notag \\
& + | \dot{\Phi}^i |^2 - r^2 | \Phi^i |^2
  - \frac{1}{3^2} | \Phi^i + i \epsilon^{ijk} J^j \circ \Phi^k |^2
  + \frac{1}{3^2} | J^i \circ \Phi^i |^2
 \notag \\
& + | \dot{\Phi}^a |^2 
  - \Big( r^2 + \frac{1}{6^2} \Big) | \Phi^a |^2
  - \frac{1}{3^2} \Phi^{a \dagger} J^i \circ ( J^i \circ \Phi^a )
 \notag \\
& - i \frac{1}{3} r 
    \Big[   \sin \left( \frac{t}{6} \right)
            ( \Phi^{0 \dagger} \Phi^4 - \Phi^{4 \dagger} \Phi^0 )
          - \cos \left( \frac{t}{6} \right)
            ( \Phi^{0 \dagger} \Phi^5 - \Phi^{5 \dagger} \Phi^0 )
    \Big] 
\bigg\} ~.
\label{lb}
\end{align}
Here, adopting the notation of \cite{Dasgupta:2002hx}, we have defined
\begin{equation}
J^i \circ M_{(rs)} \equiv J^i_{(r)} M_{(rs)} - M_{(rs)} J^i_{(s)} ~,
\label{off-com}
\end{equation}
where $M_{(rs)}$ is the $N_r \times N_s$ matrix which is a block at
$r$-th row and $s$-th column in the blocked form of a given matrix
$M$.  In the present case, $r$ and $s$ take values of $1$ and $2$.
For example, if we look at the $2 \times 2$ block matrix form of the
gauge field fluctuation $A$ in Eq.~(\ref{q-fluct}), then
$A_{(ss)}=Z^0_{(s)}$, $A_{(12)}=\Phi^0$, and $A_{(21)}=\Phi^{0
\dagger}$.

The matrix fields, $\Phi^0$, $\Phi^4$, and $\Phi^5$ are coupled with
each other through the circular motion background.  Since the time
dependent trigonometric functions may make the formulation annoying,
we consider the newly defined matrix variables as
\begin{align}
\Phi^r & \equiv   \cos \left( \frac{t}{6} \right) \Phi^4
                + \sin \left( \frac{t}{6} \right) \Phi^5 ~,
 \notag \\
\Phi^\theta &
         \equiv - \sin \left( \frac{t}{6} \right) \Phi^4
                + \cos \left( \frac{t}{6} \right) \Phi^5 ~,
\end{align}
where $\Phi^\theta$ may be interpreted as the fluctuation tangential
to the circular motion at time $t$ and $\Phi^r$ as the normal
fluctuation.  In terms of these fluctuations, the terms in the
Lagrangian (\ref{lb}), which are dependent on $\Phi^4$ and $\Phi^5$,
are rewritten as
\begin{align}
\mathrm{Tr} \bigg[ & | \dot{\Phi}^r |^2 + | \dot{\Phi}^\theta |^2
  - r^2 \left( | \Phi^r |^2 + | \Phi^\theta |^2 \right)
  - \frac{1}{3^2}
    \left(  
           \Phi^{r \dagger} J^i \circ ( J^i \circ \Phi^r )
         + \Phi^{\theta \dagger} J^i \circ ( J^i \circ \Phi^\theta ) 
    \right)
 \notag \\
& + \frac{1}{3} ( \Phi^{r \dagger} \dot{\Phi}^\theta
                   -\Phi^{\theta \dagger} \dot{\Phi}^r )
  + i \frac{r}{3}  ( \Phi^{0 \dagger} \Phi^\theta
                       -\Phi^{\theta \dagger} \Phi^0 )  \bigg] ~,
\end{align}
where we no longer see the explicit time dependent classical functions.

We are now in a position to consider the diagonalization of the
Lagrangian $L^B$.  We note that, compared to the case in the previous
section, we are in a somewhat different situation.  The fluctuation
matrices are $N_1 \times N_2$ or $N_2 \times N_1$ ones, while those in
the last section are square $N_s \times N_s$ matrices.  This means
that, when we regard an $N_1 \times N_2$ block off-diagonal matrix as
an $N_1 N_2$-dimensional reducible representation of $SU(2)$, it has
the decomposition into irreducible spin $j$ representations with the
range $|N_1-N_2|/2 \le j \le (N_1+N_2)/2-1$, that is, $N_1N_2 =
\bigoplus_{j=|N_1-N_2|/2}^{(N_1+N_2)/2-1} (2j+1)$, and may be expanded
as
\begin{align}
\Phi = \sum_{j=\frac{1}{2}|N_1-N_2|}^{\frac{1}{2}(N_1+N_2)-1} 
       \sum_{m=-j}^{j}
       \phi_{jm} Y^{N_1 \times N_2}_{jm} ~,
\label{o-exp}
\end{align}
where $Y^{N_1 \times N_2}_{jm}$ is the $N_1 \times N_2$ matrix
spherical harmonics transforming in the irreducible spin $j$
representation and $\phi_{jm}$ is the corresponding spherical mode.
The basic operation between $SU(2)$ generators $J^i_{(s)}$ and $Y^{N_1
\times N_2}_{jm}$ is given not by the commutator but by the $\circ$
operator (\ref{off-com}).  Thus the algebraic properties for the
present situation are given by Eq.~(\ref{alg}) where the commutator is
replaced by the $\circ$ operator.

Having the expansion and algebraic properties, the diagonalization
proceeds in exactly the same way as in the previous section.  Hence,
by noting that one may find a detailed procedure in
\cite{Dasgupta:2002hx}, we will present only the results of
diagonalization for the Lagrangian. We observe that, because of the
background for the circular motion in $x^4$-$x^5$ plane, the $SO(6)$
symmetry is broken to $SO(4) \times SO(2)$, while the $SO(3)$ symmetry
remains intact.  This fact naturally leads us to break the bosonic
Lagrangian (\ref{lb}) into three parts as follows:
\begin{equation}
L^B = L_{SO(3)} + L_{SO(4)} + L_\mathrm{rot}~,
\end{equation}
where $L_{SO(3)}$ is the Lagrangian for $\Phi^i$, $L_{SO(4)}$ is for
$\Phi^{a'}$ with $a'=5,6,7,8$, and $L_\mathrm{rot}$ represents the
$SO(2)$ rotational part described by $\Phi^4$, $\Phi^5$, and the gauge
fluctuation $\Phi^0$.  

We first consider $L_{SO(3)}$ and its path integration.  Its
diagonalized form is obtained by
\begin{align}
L_{SO(3)} =
&  \sum^{\frac{1}{2}(N_1+N_2)-2}_{j=\frac{1}{2}|N_1-N_2|-1}
\Bigg[  |\dot{\alpha}_{jm}|^2 
  - \left( r^2 + \frac{1}{3^2} ( j+ 1 )^2 \right)
    | \alpha_{jm} |^2 
\Bigg]
  \notag \\
&  + \sum^{\frac{1}{2}(N_1+N_2)}_{j=\frac{1}{2}|N_1-N_2|+1}
\Bigg[  | \dot{\beta}_{jm} |^2
  - \left( r^2 + \frac{1}{3^2} j^2 \right) 
    | \beta_{jm} |^2 
\Bigg]
  \notag \\
&  + \sum^{ \frac{1}{2}(N_1+N_2)-1}_{j= \frac{1}{2}|N_1-N_2| } 
\Bigg[
- | \dot{\omega}_{jm} |^2 
+ \left( r^2  + \frac{1}{3^2} j (j+1) \right)
   | \omega_{jm} |^2
\Bigg] ~,
\end{align}
where the sum of $m$ over the range $-j \le m \le j$ is implicit and
the spherical modes $\omega_{jm}$ are the degrees of freedom for the
gauge transformation, the block off-diagonal counterpart of $w_{jm}$,
(\ref{w}), in the previous section. $\alpha_{jm}$ and $\beta_{jm}$ are
the block off-diagonal counterparts of $u_{jm}$ and $v_{jm}$,
respectively.  The path integral of this Lagrangian is straightforward
and results in
\begin{align}
& \prod^{ \frac{1}{2}(N_1+N_2)-2}_{j= \frac{1}{2}|N_1-N_2|-1} 
\left[ 
 \det \left( 
          \partial_t^2 + r^2 + \frac{1}{3^2} (j+1)^2
      \right)
\right]^{-(2j+1)}
  \notag \\
\times & \prod^{\frac{1}{2}(N_1+N_2)}_{j'=\frac{1}{2}|N_1-N_2|+1}
\left[
 \det \left(
          \partial_t^2 + r^2 + \frac{1}{3^2} j'^2
      \right)
\right]^{-(2j'+1)}
  \notag \\
\times & \prod^{\frac{1}{2}(N_1+N_2)-1}_{j''=\frac{1}{2}|N_1-N_2|}
\left[
 \det \left(
          \partial_t^2 + r^2 + \frac{1}{3^2} j'' (j''+1)
      \right)
\right]^{-(2j''+1)} ~.
\label{p-ob1}
\end{align}

As for $L_{SO(4)}$, we have as its diagonalized form
\begin{equation}
L_{SO(4)} = \sum^{ \frac{1}{2}(N_1+N_2)-1}_{j= \frac{1}{2}|N_1-N_2|} 
\Bigg[
 | \dot{\phi}^{a'}_{jm} |^2 
- \left( r^2 + \frac{1}{3^2} \left( j + \frac{1}{2} \right)^2
  \right) | \phi^{a'}_{jm} |^2
\Bigg] ~,
\end{equation}
where $-j \le m \le j$ and $a'=5,6,7,8$.  Its path integration
leads us to have
\begin{equation}
\prod^{ \frac{1}{2}(N_1+N_2)-1}_{j= \frac{1}{2}|N_1-N_2|}
\left[
 \det \left(
          \partial_t^2 + r^2 
        + \frac{1}{3^2} \left( j + \frac{1}{2} \right)^2
      \right)
\right]^{- 4 (2j + 1)} ~.
\label{p-ob2}
\end{equation}

For the rotational part, the diagonalized Lagrangian is obtained as
\begin{align}
L_\mathrm{rot} =
 \sum^{ \frac{1}{2}(N_1+N_2)-1}_{j= \frac{1}{2}|N_1-N_2|}  
\Bigg[ 
&- | \dot{\phi}^0_{jm} |^2 
+ \left( r^2  + \frac{1}{3^2} j (j+1) \right)
   | \phi^0_{jm} |^2
  \notag \\
& + | \dot{\phi}^r_{jm} |^2 + | \dot{\phi}^\theta_{jm} |^2
- \left( r^2  + \frac{1}{3^2} j (j+1) \right)
  \left( | \phi^r_{jm} |^2 + | \phi^\theta_{jm} |^2 \right)
  \notag \\
& + \frac{1}{3} ( \phi^{r*}_{jm} \dot{\phi}^\theta_{jm}
                   -\phi^{\theta *}_{jm} \dot{\phi}^r_{jm} )
  + i \frac{r}{3}
       ( \phi^{0*}_{jm} \phi^\theta_{jm}
        -\phi^{\theta *}_{jm} \phi^0_{jm} ) \Bigg] ~,
\end{align}
where $-j \le m \le j$.  Since all the coupling coefficients between
the modes are time-independent constants,  the path integral of this
Lagrangian is readily evaluated and gives
\begin{align}
\prod^{ \frac{1}{2}(N_1+N_2)-1}_{j= \frac{1}{2}|N_1-N_2|} 
&\Bigg[
 \det \left(
          \partial_t^2 + r^2 + \frac{1}{3^2} j (j+1)
      \right)
 \det \left(
          \partial_t^2 + r^2 + \frac{1}{3^2} j^2
      \right)
 \notag \\
& \times  \det \left(
          \partial_t^2 + r^2 + \frac{1}{3^2} (j+1)^2
      \right)
\Bigg]^{-(2j + 1)} ~.
\label{p-ob3}
\end{align}

\subsection{Fermionic fluctuation}

The fermionic Lagrangian of the block off-diagonal action
$S_\text{off-diag}$, (\ref{oda}), is
\begin{align}
L^F =
 2 \mathrm{Tr}
\left[ \,
     i \chi^\dagger \dot{\chi}
     - i \frac{1}{4} \chi^\dagger \gamma^{123} \chi
     + \frac{1}{3} \chi^\dagger \gamma^i J^i \circ \chi
     + r \chi^\dagger
        \left(  \gamma^4 \cos \left( \frac{t}{6} \right)
              + \gamma^5 \sin \left( \frac{t}{6} \right)
        \right)  \chi \,
\right] \,,
\label{lf1}
\end{align}
As was done in Eq.~(\ref{s-dec}), we decompose the fermion $\chi$ into
$\chi_{A\alpha}$ and $\hat{\chi}^{A\beta}$ according to ${\bf 16}
\rightarrow ({\bf 2}, {\bf 4}) + (\bar{\bf 2}, \bar{\bf 4})$.
Then the expression of $\chi$ may be taken as
\begin{equation}
\chi = \frac{1}{\sqrt{2}}
\begin{pmatrix}
\chi_{A\alpha} \\ 
\hat{\chi}^A_\alpha
\end{pmatrix} ~,
\label{o-dec}
\end{equation}
where $\hat{\chi}^A_\alpha=\epsilon_{\alpha\beta}
\hat{\chi}^{A\beta}$.  It should be noted that, contrary to the block
diagonal fermionic matrix $\Psi_{(s)}$ of (\ref{d-dec}), $\chi$ is the
two copy of the $SO(9)$ representation ${\bf 16}$ as one may see from
Eq.~(\ref{q-fluct}).  This means that $\chi_{A\alpha}$ and
$\hat{\chi}^A_\alpha$ should be treated as independent spinors not
related in any way.  By plugging the decomposition (\ref{o-dec}) into
the Lagrangian (\ref{lf1}), we have
\begin{align}
L^F = \mathrm{Tr}
\Bigg[
&   i \chi^{\dagger A\alpha} \dot{\chi}_{A\alpha} 
  - \frac{1}{4} \chi^{\dagger A\alpha} \chi_{A\alpha}
  - \frac{1}{3} \chi^{\dagger A\alpha}(\sigma^i)_{\alpha}{}^\beta
                J^i\circ \chi_{A\beta} 
\notag \\
& + i \hat{\chi}^{\dagger \alpha}_A \dot{\hat{\chi}}^A_\alpha
  + \frac{1}{4} \hat{\chi}^{\dagger \alpha}_A \hat{\chi}^A_\alpha
  + \frac{1}{3} \hat{\chi}^{\dagger \alpha}_A 
                (\sigma^i)_{\alpha}{}^\beta
                J^i\circ \hat{\chi}^A_\beta
\notag \\
& + r \chi^{\dagger A\alpha}
      \left(
            \rho^4_{AB} \cos \left( \frac{t}{6} \right)
          + \rho^5_{AB} \sin \left( \frac{t}{6} \right)
      \right) \hat{\chi}^B_\alpha 
\notag \\
& + r \hat{\chi}^{\dagger \alpha}_A
      \left(
            (\rho^4)^{\dagger AB} \cos \left( \frac{t}{6} \right)
          + (\rho^5)^{\dagger AB} \sin \left( \frac{t}{6} \right)
      \right) \chi_{B \alpha}
\Bigg] \, .
\label{lf2}
\end{align}
The Lagrangian has explicit time dependence due to the presence of
the circular motion background.  In order to hide it, we take the
fermionic field $\hat{\chi}^A_\alpha$ as
\begin{align}
\hat{\chi}^A_\alpha \equiv
  \left(
            (\rho^4)^{\dagger AB} \cos \left( \frac{t}{6} \right)
          + (\rho^5)^{\dagger AB} \sin \left( \frac{t}{6} \right)
  \right)  \tilde{\chi}_{B \alpha} ~,
\end{align}
where we have introduced a new fermionic field $\tilde{\chi}_{A
\alpha}$ which is in the ${\bf 4}$ of $SU(4)$.  Then, by using
the following identities,
\begin{align}
&\left(
            \rho^4 \cos \left( \frac{t}{6} \right)
          + \rho^5 \sin \left( \frac{t}{6} \right)
 \right)
 \left(
            (\rho^4)^\dagger \cos \left( \frac{t}{6} \right)
          + (\rho^5)^\dagger \sin \left( \frac{t}{6} \right)
 \right)
= 1 ~, 
\notag \\
&\left(
            \rho^4 \cos \left( \frac{t}{6} \right)
          + \rho^5 \sin \left( \frac{t}{6} \right)
      \right)
 \left(
          - (\rho^4)^\dagger \sin \left( \frac{t}{6} \right)
          + (\rho^5)^\dagger \cos \left( \frac{t}{6} \right)
      \right)
= \rho^4 (\rho^5)^\dagger ~,
\end{align}
which are proved via the Clifford algebra (\ref{cliff}), we
may show that the Lagrangian (\ref{lf2}) becomes
\begin{align}
L^F = \mathrm{Tr}
\bigg[
& i \chi^{\dagger A\alpha} \dot{\chi}_{A\alpha} 
  - \frac{1}{4} \chi^{\dagger A\alpha} \chi_{A\alpha}
  - \frac{1}{3} \chi^{\dagger A\alpha}(\sigma^i)_{\alpha}{}^\beta
                J^i\circ \chi_{A\beta} 
\notag \\
& + i\tilde{\chi}^{\dagger A\alpha} \dot{\tilde{\chi}}_{A\alpha} 
  + \frac{1}{4} \tilde{\chi}^{\dagger A\alpha} 
                \tilde{\chi}_{A\alpha}
  + \frac{1}{3} \tilde{\chi}^{\dagger A\alpha}
                (\sigma^i)_{\alpha}{}^\beta
                J^i\circ \tilde{\chi}_{A\beta}
\notag \\
& + r \left( \chi^{\dagger A \alpha} \tilde{\chi}_{A \alpha} 
            +\tilde{\chi}^{\dagger A \alpha} \chi_{A \alpha}
      \right)
  + \frac{i}{6} \tilde{\chi}^{\dagger A\alpha}
                ( \rho^4 (\rho^5)^\dagger )_A{}^B 
                \tilde{\chi}_{B\alpha}
\bigg] \, .
\label{lf3}
\end{align}
The explicit time dependent classical functions disappear and the term
containing $\rho^4 (\rho^5)^\dagger$ appears, which originates from
the kinetic term of $\hat{\chi}^A_\alpha$ in (\ref{lf2}).

With the above Lagrangian (\ref{lf3}), the diagonalization proceeds in
the same manner with that for the block diagonal fermionic Lagrangian
(\ref{diag-lf}), except for some differences pointed out in the
previous subsection.  In the expansion of $\chi_\alpha$ and
$\tilde{\chi}_\alpha$ ($SU(4)$ indices are suppressed.) in terms of
the matrix spherical harmonics like (\ref{o-exp}), let us denote their
spherical modes as $(\chi_\alpha)_{jm}$ and
$(\tilde{\chi}_\alpha)_{jm}$ respectively.  Then the diagonalization
results in $(\chi_\alpha)_{jm} \rightarrow ( \pi_{jm}, \eta_{jm})$ and
$(\tilde{\chi}_\alpha)_{jm} \rightarrow ( \tilde{\pi}_{jm},
\tilde{\eta}_{jm})$.  The modes $\pi_{jm}$ and $\tilde{\pi}_{jm}$ have
the same mass of $\frac{1}{3} (j+\frac{3}{4})$ with $\frac{1}{2}|N_1 -
N_2| - \frac{1}{2} \le j \le\frac{1}{2}(N_1 + N_2)-\frac{3}{2}$.  For
the modes $\eta_{jm}$ and $\tilde{\eta}_{jm}$, their mass is
$\frac{1}{3} (j+\frac{1}{4})$ with $\frac{1}{2}|N_1 - N_2|+\frac{1}{2}
\le j \le\frac{1}{2}(N_1 + N_2)-\frac{1}{2}$.  All the modes have the
same range of $m$ as $-j \le m \le j$.  With respect to these
diagonalized spherical modes, the Lagrangian (\ref{lf3}) is written as
\begin{align}
L^F = 
 \sum_{j=\frac{1}{2}|N_1 - N_2| - \frac{1}{2}}^{\frac{1}{2}(N_1
        + N_2)-\frac{3}{2}} 
\Bigg[ \, 
&
    i \pi_{jm}^\dagger \dot{\pi}_{jm} 
  + i \tilde{\pi}_{jm}^\dagger \dot{\tilde{\pi}}_{jm} 
    - \frac{1}{3} \left(j + \frac{3}{4} \right) 
         (  \pi^\dagger_{jm} \pi_{jm} 
          - \tilde{\pi}^\dagger_{jm} \tilde{\pi}_{jm} )
\notag \\ 
& + r (  \pi^\dagger_{jm} \tilde{\pi}_{jm}
       + \tilde{\pi}^\dagger_{jm} \pi_{jm} )
  + \frac{i}{6} \tilde{\pi}^\dagger_{jm} 
                \rho^4 (\rho^5)^\dagger \tilde{\pi}_{jm} \,
\Bigg]
\notag \\ 
+\sum_{j=\frac{1}{2}|N_1 - N_2| + \frac{1}{2}}^{\frac{1}{2}(N_1
        + N_2)-\frac{1}{2}} 
\Bigg[ \, 
&
    i \eta_{jm}^\dagger \dot{\eta}_{jm} 
  + i \tilde{\eta}_{jm}^\dagger \dot{\tilde{\eta}}_{jm} 
    - \frac{1}{3} \left(j + \frac{1}{4} \right) 
         (  \eta^\dagger_{jm} \eta_{jm}
          - \tilde{\eta}^\dagger_{jm} \tilde{\eta}_{jm} )
\notag \\ 
& + r (  \eta^\dagger_{jm} \tilde{\eta}_{jm}
       + \tilde{\eta}^\dagger_{jm} \eta_{jm} )
  + \frac{i}{6} \tilde{\eta}^\dagger_{jm} 
                \rho^4 (\rho^5)^\dagger \tilde{\eta}_{jm} \,
\Bigg] ~,
\label{lf4}
\end{align}
where $-j \le m \le j$ and the $SU(4)$ indices are suppressed.

The product $\rho^4 (\rho^5)^\dagger$ measures the $SO(2)$ chirality
in the $x^4$-$x^5$ plane where the circular motion takes place.  Since
$(\rho^4 (\rho^5)^\dagger)^2=-1$, its eigenvalues are $\pm i$.  Each
spherical mode may split into modes having definite $\rho^4
(\rho^5)^\dagger$ eigenvalues as follows:
\begin{align}
&\pi_{jm} = \pi_{+jm} + \pi_{-jm} ~,
&& \eta_{jm} = \eta_{+jm} + \eta_{-jm} ~,
\notag \\
&\tilde{\pi}_{jm}=\tilde{\pi}_{+jm}+\tilde{\pi}_{-jm} ~,
&& \tilde{\eta}_{jm}=\tilde{\eta}_{+jm}+\tilde{\eta}_{-jm} ~,
\end{align}
where the modes on the right hand sides satisfy
\begin{align}
& \rho^4 (\rho^5)^\dagger \pi_{\pm jm} 
              = \pm i \pi_{\pm jm}~,
& \rho^4 (\rho^5)^\dagger \tilde{\pi}_{\pm jm} 
              = \pm i \tilde{\pi}_{\pm jm}~,
\notag \\
& \rho^4 (\rho^5)^\dagger \eta_{\pm jm} 
              = \pm i \eta_{\pm jm}~,
& \rho^4 (\rho^5)^\dagger \tilde{\eta}_{\pm jm} 
              = \pm i \tilde{\eta}_{\pm jm}~.
\end{align}
One may see that the Lagrangian (\ref{lf4}) composed of two
independent parts.  One is for $\pi_{jm}$ and $\tilde{\pi}_{jm}$, and
the other one for $\eta_{jm}$ and $\tilde{\eta}_{jm}$.  If we first
consider the part for $\pi_{jm}$ and $\tilde{\pi}_{jm}$, then,
according to the above splitting of modes, we have
\begin{align}
\sum_{j=\frac{1}{2}|N_1 - N_2| - \frac{1}{2}}^{\frac{1}{2}(N_1
        + N_2)-\frac{3}{2}} 
\Bigg[ \, 
&   i \pi_{+jm}^\dagger \dot{\pi}_{+jm} 
  + i \pi_{-jm}^\dagger \dot{\pi}_{-jm}
  - \frac{1}{3} \left(j + \frac{3}{4} \right) 
      (  \pi^\dagger_{+jm} \pi_{+jm} 
       + \pi^\dagger_{-jm} \pi_{-jm} )
\notag \\
&  + i \tilde{\pi}_{+jm}^\dagger \dot{\tilde{\pi}}_{+jm} 
   + \frac{1}{3} \left(j + \frac{1}{4} \right) 
       \tilde{\pi}^\dagger_{+jm} \tilde{\pi}_{+jm}
   + i \tilde{\pi}_{-jm}^\dagger \dot{\tilde{\pi}}_{-jm} 
   + \frac{1}{3} \left(j + \frac{5}{4} \right) 
       \tilde{\pi}^\dagger_{-jm} \tilde{\pi}_{-jm} 
\notag \\
&  + r (  \pi^\dagger_{+jm} \tilde{\pi}_{+jm}
        + \tilde{\pi}^\dagger_{+jm} \pi_{+jm} )
   + r (  \pi^\dagger_{-jm} \tilde{\pi}_{-jm}
        + \tilde{\pi}^\dagger_{-jm} \pi_{-jm}  ) \,
\Bigg] ~,
\end{align}
where again $-j \le m \le j$.  The path integration of this part is
straightforward and gives
\begin{align}
\prod_{j=\frac{1}{2}|N_1 - N_2| - \frac{1}{2}}^{\frac{1}{2}(N_1
        + N_2)-\frac{3}{2}} 
\Bigg[ \,
&
\det \left( \partial_t^2 + \frac{i}{6} \partial_t + r^2 
          + \frac{1}{3^2} \left(j + \frac{1}{2} \right)^2
          - \frac{1}{3^2 \cdot 4^2}
     \right)
\notag \\
\times &
\det \left( \partial_t^2 - \frac{i}{6} \partial_t + r^2 
          + \frac{1}{3^2} \left(j + \frac{1}{2} \right)^2
          - \frac{1}{3^2 \cdot 4^2}
     \right)
\notag \\
\times &
\det \left( \partial_t^2 + \frac{i}{6} \partial_t + r^2 
          + \frac{1}{3^2} (j + 1)^2
          - \frac{1}{3^2 \cdot 4^2}
     \right)
\notag \\
\times &
\det \left( \partial_t^2 - \frac{i}{6} \partial_t + r^2 
          + \frac{1}{3^2} (j + 1 )^2
          - \frac{1}{3^2 \cdot 4^2}
     \right) \,
\Bigg]^{2j+1} ~.
\label{p-of1}
\end{align}
The other part for $\eta_{jm}$ and $\tilde{\eta}_{jm}$ is obtained as
\begin{align}
\sum_{j=\frac{1}{2}|N_1 - N_2| + \frac{1}{2}}^{\frac{1}{2}(N_1
        + N_2)-\frac{1}{2}} 
\Bigg[ \, 
&   i \eta_{+jm}^\dagger \dot{\eta}_{+jm} 
  + i \eta_{-jm}^\dagger \dot{\eta}_{-jm}
  - \frac{1}{3} \left(j + \frac{1}{4} \right) 
      (  \eta^\dagger_{+jm} \eta_{+jm} 
       + \eta^\dagger_{-jm} \eta_{-jm} )
\notag \\
&  + i \tilde{\eta}_{+jm}^\dagger \dot{\tilde{\eta}}_{+jm} 
   + \frac{1}{3} \left(j - \frac{1}{4} \right) 
       \tilde{\eta}^\dagger_{+jm} \tilde{\eta}_{+jm}
   + i \tilde{\eta}_{-jm}^\dagger \dot{\tilde{\eta}}_{-jm} 
   +  \frac{1}{3} \left(j + \frac{3}{4} \right) 
       \tilde{\eta}^\dagger_{-jm} \tilde{\eta}_{-jm} 
\notag \\
&   + r (  \eta^\dagger_{+jm} \tilde{\eta}_{+jm}
        + \tilde{\eta}^\dagger_{+jm} \eta_{+jm} ) 
    + r (  \eta^\dagger_{-jm} \tilde{\eta}_{-jm}
        + \tilde{\eta}^\dagger_{-jm} \eta_{-jm} ) \,
\Bigg] ~,
\end{align}
where the sum over $m$ for $-j \le m \le j$ is implicit.  The
path integration of this part results in
\begin{align}
\prod_{j=\frac{1}{2}|N_1 - N_2| + \frac{1}{2}}^{\frac{1}{2}(N_1
        + N_2)-\frac{1}{2}} 
\Bigg[ \, 
&
\det \left( \partial_t^2 + \frac{i}{6} \partial_t + r^2 
          + \frac{1}{3^2} j^2
          - \frac{1}{3^2 \cdot 4^2}
     \right)
\notag \\
\times &
\det \left( \partial_t^2 - \frac{i}{6} \partial_t + r^2 
          + \frac{1}{3^2} j^2
          - \frac{1}{3^2 \cdot 4^2}
     \right) \,
\notag \\
\times &
\det \left( \partial_t^2 + \frac{i}{6} \partial_t + r^2 
          + \frac{1}{3^2} \left(j + \frac{1}{2} \right)^2
          - \frac{1}{3^2 \cdot 4^2}
     \right)
\notag \\
\times &
\det \left( \partial_t^2 - \frac{i}{6} \partial_t + r^2 
          + \frac{1}{3^2} \left(j + \frac{1}{2} \right)^2
          - \frac{1}{3^2 \cdot 4^2}
     \right) \,
\Bigg]^{2j+1} ~.
\label{p-of2}
\end{align}

\subsection{Ghost fluctuation}

Finally, we consider the path integration for the ghost part of the
action $S_\text{off-diag}$ (\ref{oda}).  The block off-diagonal
Lagrangian for the ghosts is
\begin{align}
L^G =  \mathrm{Tr}
\bigg[ \, 
& \dot{\bar{C}} \dot{C}^{\dagger} 
+ r^2 \bar{C} C^{\dagger}
+ \frac{1}{3^2} (J^i\circ \bar{C})(J^i\circ C^\dagger) \,
\bigg]  
\notag \\
+ \mathrm{Tr} 
\bigg[ \,
& \dot{\bar{C}}^\dagger \dot{C} 
+ r^2 \bar{C}^\dagger C
+ \frac{1}{3^2} (J^i\circ \bar{C}^\dagger)(J^i\circ C) \,
\bigg] \, . 
\end{align}
The diagonalization may be carried out by using the same logic
in the previous subsections.  We expand the ghost fields
$C$ and $\bar{C}$ in terms of the matrix spherical harmonics
according to (\ref{o-exp}), and denote their spherical modes
as $c_{jm}$ and $\bar{c}_{jm}$ respectively. Then the
diagonalized Lagrangian is obtained as 
\begin{align}
L^G = \sum^{ \frac{1}{2}(N_1+N_2)-1}_{j= \frac{1}{2}|N_1-N_2|} 
\Bigg[ \,
  \dot{\bar{c}}^*_{jm} \dot{c}^{\dagger}_{jm} 
+ \dot{\bar{c}}^{\dagger *}_{jm} \dot{c}_{jm} 
  - \left( r^2 + \frac{1}{3^2} j(j+1) \right)
  ( \bar{c}^*_{jm} c^{\dagger}_{jm} 
  + \bar{c}^{\dagger *}_{jm} c_{jm} ) \,
\Bigg]\,,
\end{align}
where the sum over $m$ for the range $-j \le m \le j$ is implicit.

The path integral for the above diagonalized Lagrangian is then
immediately evaluated as follows:
\begin{equation}
\prod^{ \frac{1}{2}(N_1+N_2)-1}_{j= \frac{1}{2}|N_1-N_2|} 
\bigg[
 \det \left(
          \partial_t^2 + r^2 + \frac{1}{3^2} j (j+1)
      \right)
\bigg]^{2 (2j + 1)} ~.
\label{p-og}
\end{equation}

\subsection{Effective potential}

Having evaluated the path integral for each part of the block
off-diagonal action $S_\text{off-diag}$ (\ref{oda}), the one-loop
effective action, $\Gamma_\text{eff}^\text{(int)}$, which describes
the interaction between two membrane fuzzy spheres with the classical
configuration (\ref{b3}) and (\ref{b6}), is now given by
\begin{align}
e^{i \Gamma_\text{eff}^\text{(int)} }
= [(\ref{p-ob1}) \times (\ref{p-ob2}) \times (\ref{p-ob3})]_B
 \times [(\ref{p-of1}) \times (\ref{p-of2})]_F
 \times [(\ref{p-og})]_G ~,
\end{align}
where the subscripts on the right hand side denote the bosonic, the
fermionic, and the ghost contributions.  We see that the ghost
contribution, Eq.~(\ref{p-og}), eliminate those of unphysical gauge
degrees of freedom present in bosonic contributions,
Eqs.~(\ref{p-ob1}) and (\ref{p-ob3}).  Thus only the physical degrees
of freedom contribute to the effective action, as it should be.

The explicit expression of the effective action is obtained by
consulting the generic result presented from Eq.~(\ref{gen}) to
(\ref{gen-eff}).  The effective potential, $V_\text{eff}$, about which
we are concerned in this subsection, is then given by
$\Gamma_\text{eff}^\text{(int)} = - \int dt V_\text{eff}$.  In
expressing the effective potential, it is convenient to write
$V_\text{eff}$ as
\begin{align}
V_\text{eff} = V_\text{eff}^B + V_\text{eff}^F ~,
\label{ve}
\end{align}
where $V_\text{eff}^B$ ($V_\text{eff}^F$) is the contribution of the
physical bosonic (fermionic) degrees of freedom to the effective
potential.  The expression that we obtain for $V_\text{eff}^B$ is then
\begin{align}
V^B_\mathrm{eff} =
&  \sum^{ \frac{1}{2}(N_1+N_2)-2}_{j= \frac{1}{2}|N_1-N_2|-1} (2j+1)
      \sqrt{ r^2 + \frac{1}{3^2} (j+1)^2 }
  +\sum^{\frac{1}{2}(N_1+N_2)}_{j=\frac{1}{2}|N_1-N_2|+1} (2j+1)
      \sqrt{ r^2 + \frac{1}{3^2} j^2 }
  \notag \\
& + \sum^{\frac{1}{2}(N_1+N_2)-1}_{j=\frac{1}{2}|N_1-N_2|} 4 (2j+1)
       \sqrt{ r^2 + \frac{1}{3^2}
             \left( j + \frac{1}{2} \right)^2 }
  \notag \\
& + \sum^{\frac{1}{2}(N_1+N_2)-1}_{j=\frac{1}{2}|N_1-N_2|} (2j+1)
   \Bigg[ \;
       \sqrt{ r^2 + \frac{1}{3^2} (j+1)^2 }
     + \sqrt{ r^2 + \frac{1}{3^2} j^2 } \;
   \Bigg] ~,
\end{align}
while, for the fermionic contribution $V_\text{eff}^F$, we obtain
\begin{align}
V_{\rm eff}^F =
& - \sum_{j=\frac{1}{2}|N_1 - N_2| 
        - \frac{1}{2}}^{\frac{1}{2}(N_1 + N_2) - \frac{3}{2}}
  2 (2j+1) 
 \left[ \sqrt{r^2 + \frac{1}{3^2} (j + 1)^2}
       +\sqrt{r^2 + \frac{1}{3^2} 
               \left(j + \frac{1}{2}\right)^2 } \;
 \right]
\nonumber \\
& -\sum_{j=\frac{1}{2}|N_1 - N_2| 
         + \frac{1}{2}}^{\frac{1}{2}(N_1 + N_2) - \frac{1}{2}} 
  2 (2j+1) 
 \left[ \sqrt{r^2 + \frac{1}{3^2}
                \left(j + \frac{1}{2}\right)^2 } \,
       +\sqrt{r^2 + \frac{1}{3^2} j^2} \;
 \right] \, . 
\end{align}

The above expressions show that $V_\text{eff}^B$ and $V_\text{eff}^F$
have the same structure except for the ranges of $j$.  This leads us
to expect a great amount of cancellation.  Indeed, what we have found
is that they are exactly the same.  One way to see the cancellation is
to adjust all the ranges of the summation parameter $j$ to the range
$\frac{1}{2}|N_1-N_2| \le j \le \frac{1}{2}(N_1+N_2)-1$.  Therefore,
the one-loop effective potential $V_\text{eff}$, (\ref{ve}), as a
function of the distance $r$ between two fuzzy spheres is just flat
potential;
\begin{align}
V_\text{eff} (r) = 0 ~.
\label{veff}
\end{align}

\section{Conclusion and discussion}
\label{conc}

We have studied the one-loop quantum corrections to a classical
background of the plane-wave matrix model in the framework of the path
integration.  The background is composed of two supersymmetric
membrane fuzzy spheres, (\ref{b3}) and (\ref{b6}).  One fuzzy sphere
is located at the origin in the $SO(6)$ symmetric space and the other
one rotates around it with the distance $r$.

Firstly, the quantum stability of each fuzzy sphere has been shown.
In fact, the stability check already has been done in the operator
\cite{Dasgupta:2002hx} as well as in the path integral
\cite{Sugiyama:2002bw} formulation.  However, while the size $N_s$ of
the fuzzy sphere has been taken arbitrary in the operator formulation,
it has been restricted to the minimal one, that is $N_s=2$, in the path
integral formulation.  Since the fuzzy sphere in this paper has an
arbitrary size, our stability check can be regarded as the full
generalization of the previous path integral result.

Secondly, the one-loop effective potential describing the interaction
between two fuzzy spheres has been calculated as a function of the
distance $r$.  Interestingly, the result is that the effective
potential $V_\text{eff}$ is flat and thus the fuzzy spheres do not
feel any force.  This implies that the whole configuration of two
fuzzy spheres given by Eqs.~(\ref{b3}) and (\ref{b6}) is
supersymmetric.  Although the flatness of the effective potential is
the one-loop result, we expect that the result holds also for higher
loops.  For the supersymmetric properties of the fuzzy sphere
configuration itself, the study of supersymmetry algebra may be more
helpful rather than the path integral formulation.  It would be
interesting to investigate the configuration in this paper through the
supersymmetry algebra.

Let us consider the radial distance $r$ between two fuzzy spheres and
discuss about its possible interpretation.  We first consider the form
of the effective action when $r$ is time dependent.  Since the
circular motion is taken as the background, $r$ is constant in this
paper.  However, if we slightly deform the circular motion to the
elliptic one, $r$ can be made to have time dependence.  We can also
make the time variation of it, $\dot{r}$, arbitrarily small by
controlling the degree of deformation.  In this case, it is expected
that the fuzzy spheres begin to interact and the effective action
$\Gamma_\text{eff}^\text{(int)}$ may be written as
\begin{align}
\Gamma_\text{eff}^\text{(int)} 
= \frac{\mu^3 N_1}{2} \int dt \, \dot{r}^2 
  + f( \dot{r},r ) + \mathcal{O} (\mu^{-3/2}) ~,
\end{align}
where the kinetic term for $r$ is the value of the classical action
$S_0$ with the rescaling (\ref{rescale}), $f( \dot{r}, r)$ is the
would-be one-loop contribution to the effective action with the
property $f(0,r)=0$, and the term of order $\mu^{-3/2}$ implies the
higher loop corrections.  

If $\dot{r}=0$, the above effective action vanishes and the
supersymmetric situation is recovered.  This is reminiscent of the
effective action for graviton-graviton scattering in the flat space
matrix model \cite{Banks:1997vh,Becker:1997xw}.  The distance between
two gravitons comes from the flat directions which are continuous
moduli or supersymmetric vacua making the potential of the flat space
matrix model vanish.  In the plane-wave matrix model, it is known that
there is no continuous moduli and hence we do not have flat
directions.  However, if we look at the tree level action (\ref{clav})
evaluated for the fuzzy sphere configuration, (\ref{b3}) and
(\ref{b6}), we see that it vanishes exactly and does not depend on $r$
which is continuous from $0$ to $\infty$.  This means that, as long as
the fuzzy sphere dynamics is concerned, the radius of the circular
motion may be interpreted as the flat direction.  

In this paper, the circular motion takes place in the $x^4$-$x^5$
sub-plane of the $SO(6)$ symmetric space.  Since the $SO(6)$ symmetric
space has two other sub-planes, that is $x^6$-$x^7$ and $x^8$-$x^9$,
and we may embed the circular motion in one of those, there may be
three flat directions in total, which are radial directions of three
sub-planes.  However, it is not obvious whether these three directions
are connected in a continuous way or not, because of the
supersymmetric property of rotating fuzzy sphere; all the points in
the supersymmetric moduli are expected to preserve a fixed fraction of
supersymmetry.  While the fuzzy sphere rotating in only one sub-plane
is 1/2-BPS object, it is generically 1/4-BPS when it has angular
momenta also in the other sub-planes \cite{Park:2002cb}.  If we turn
to the $SO(3)$ symmetric space, it seems that we do not have flat
directions, since only the fuzzy sphere rotating with fixed radius is
supersymmetric \cite{Bak:2002rq,Mikhailov:2002wx,Park:2002cb}.

In view of the gauge/gravity duality, it is interesting to study the
same situation in the supergravity side.  In the large $N$ limit, the
leading order interaction terms obtained from the supergravity side
analysis would match with those from the matrix theory analysis.  We
expect that the supergravity side analysis is helpful and provides
clearer understanding about the structure of the effective potential.

The calculation in this paper has been carried out in the Minkowskian
time signature.  The basic reason why we have not taken the Wick
rotation for some convenience in the actual calculation is the
periodic nature of the circular motion background (\ref{b6}).  The
naive change of the Minkowskian time to the Euclidean one in the
process of calculation breaks the periodicity of the background and
the reality of the action.  If one wants to study the time dependent
background with the Euclidean time signature, he or she should begin
with the Euclidean action at the first setup.  In the Euclidean case,
the time dependent solutions to the equations of motion are given by
hyperbolic functions, which give open paths not periodic ones.  In the
sense that there is no classical background solution corresponding to
open path in the Minkowskian time plane-wave matrix model, it would be
interesting to consider the model in the Euclidean time for studying
such a background.

In the situation where the one-loop effective potential is flat, it is
a natural step to consider the case where the radius $r$ between two
fuzzy spheres is time dependent and calculate the form of the
interaction.  We hope to return to this issue in the near future.

\section*{Acknowledgments}
One of us, K.Y., would like to thank M.~Sakaguchi and D.~Tomino for
helpful discussions and comments.  The work of H.S. was supported by
Korea Research Foundation (KRF) Grant KRF-2001-015-DP0082.

\end{document}